\documentclass[aps,prl,reprint,superscriptaddress]{revtex4-2}
\usepackage{amsmath,amssymb}
\usepackage{graphicx}
\usepackage[dvipsnames]{xcolor}
\usepackage[T1]{fontenc}
\usepackage{physics}
\usepackage{hyperref}
\usepackage{bm}
\usepackage[english]{babel}

\begin{document}

\title{Memory-aware acceleration of orientational dynamics in nanoparticle suspensions}

\author{Miguel Ib\'a\~nez}
\affiliation{Universidad de Granada, Department of Applied Physics and Research Unit ‘Modeling Nature’ (MNat), Nanoparticle Trapping Laboratory, Granada, Spain}
\author{Ra\'ul A. Rica-Alarc\'on}
\email[Corresponding author; email: ]{rul@ugr.es}
\affiliation{Universidad de Granada, Department of Applied Physics and Research Unit ‘Modeling Nature’ (MNat), Nanoparticle Trapping Laboratory, Granada, Spain}

\author{Mar\'ia L. Jim\'enez}
\affiliation{Universidad de Granada, Department of Applied Physics and Research Unit ‘Modeling Nature’ (MNat), Nanoparticle Trapping Laboratory, Granada, Spain}

\begin{abstract}

The relaxation of stochastic systems after sudden perturbations is constrained by speed limits and often reveals memory effects that hinder attempts to accelerate their dynamics. Here we demonstrate Kovacs-type nonmonotonic relaxation in the electro-orientation of non-spherical nanoparticles and show how this memory effect limits simple acceleration protocols. Experimentally, the orientational dynamics is monitored optically through field-induced birefringence, which is proportional to the nematic order parameter. When an AC electric field is first set to an extreme value until the birefringence reaches its target and is then switched to the target field (matched two-step protocol), the relaxation exhibits a characteristic Kovacs shoulder. We interpret this behavior within a theoretical framework based on the Smoluchowski equation for the orientational probability density. In the high-frequency AC regime, orientational relaxation is governed by induced dipoles, and the observed memory effect originates from polydispersity, which generates a spectrum of rotational diffusion coefficients and hence multiscale relaxation. Building on this insight, we design protocols that mitigate the detrimental effect of memory by sequentially suppressing the slowest active relaxation mode. Experiments on nanoparticle suspensions with different properties confirm these mechanisms, and we demonstrate substantial reductions in relaxation time compared with single quenches and matched two-step protocols with NaMt suspensions. More broadly, these results illustrate how memory effects emerge when many degrees of freedom are steered with a single control parameter and provide an experimentally accessible strategy for controlling multiscale stochastic dynamics.

\end{abstract}

\maketitle

The response of a system to a sudden change in a control parameter often follows an exponential relaxation set by its natural time scale. Whether imposed by application demands or simply by the desire to accelerate such transients, a variety of engineered protocols have been proposed to minimize duration~\cite{martinez2016engineered,frim2021engineered,chupeau2018engineered,rengifo2024analytical,pemartin2024shortcuts}, energy~\cite{monter2025optimal} or to optimize trade-offs between time and entropy production~\cite{pires2023optimal}. These trade-offs are tightly linked to stochastic speed limits in classical systems~\cite{shiraishi2018speed,ito2020stochastic,guery2023driving}. Intriguingly, the search for minimal-time protocols—such as strategies inspired in the brachistochrone problem~\cite{patron2022thermal,ibanez2026thermal}—often exposes anomalies: the nonmonotonic relaxation seen in two-step protocols (the Kovacs effect)~\cite{kovacs1979isobaric,prados2010kovacs,prados2014kovacs,lahini2017nonmonotonic,kursten2017giant,dillavou2018nonmonotonic,murphy2020memory,morgan2020glassy,militaru2021kovacs,tong2023strain,patron2023non}, the Mpemba effect, where hotter initial states may relax faster to a cold state than a mild one~\cite{lasanta2017hotter,kumar2020exponentially,kumar2022anomalous,patron2023non,biswas2023mpemba,teza2023eigenvalue,teza2025speedups}, and heating/cooling time asymmetries under harmonic confinement~\cite{lapolla2020faster,ibanez2024heating,meibohm2021relaxation,van2021toward,dieball2023asymmetric}. Such behaviors show that out-of-equilibrium evolution cannot be reduced to a sequence of equilibria and that memory may arise broadly~\cite{keim2019memory}.

Orientational control of non-spherical nanoparticles dispersed in a liquid alters the macroscopic properties of the system—including its mechanical response~\cite{ramos2009electrorheology}, and optical behavior, such as birefringence~\cite{ArenasGuerrero2016,ArenasGuerrero2021,Li2022,Dong2024}. Optical birefringence is widely used to probe orientational order in soft-matter systems ranging from liquid-crystal droplets and shells to anisotropic colloids~\cite{mach2002electro,Mantegazza2005,cazorla2024electro}. Precise alignment within conducting films and thin-film transistors enables high-performance displays, flexible electronics, and photovoltaics~\cite{Xi2024}, while the giant birefringence of aligned two-dimensional nanosheets is attractive for solar-blind ultraviolet communications~\cite{Ding2020,Xu2024}. Control over birefringence underpins metamaterials, sensing, and data encryption~\cite{Khan2024,Wang2022}, as well as optical components such as liquid-crystal fibers for polarization modulation~\cite{Budaszewski2018,Gyaprasad2024}.

Electric fields can induce orientation of anisotropic particles immersed in a liquid medium~\cite{arcenegui2013electro} by exerting torques that favor alignment along the field. The characteristic alignment time reflects competition between field-induced and Brownian torques; minimizing it is central to both fundamental studies and scale-up applications. In situations where several alignment mechanisms coexist, the transient response can exhibit multiple time scales and nontrivial dynamics. Non-equilibrium thermodynamics provides a framework for describing and steering these transients, yet general time-optimal protocols remain elusive~\cite{guery2023driving}.

Here we show experimentally that the transient electro-orientation following a sudden change in field amplitude, probed via optical birefringence, exhibits the nonmonotonic relaxation characteristic of the Kovacs effect, arising from the coexistence of multiple relaxation times due to polydispersity. We provide a theoretical explanation based on the Smoluchowski description of orientational dynamics in monodisperse ensembles and identify conditions that isolate a single effective time scale. Leveraging this structure, we design and implement conceptually clean, experimentally feasible protocols that accelerate transients and connect non-equilibrium steady states in times substantially shorter than those set by natural exponential relaxation. 

\section{Transient behavior in the electro-orientation of non-spherical particles}

When an electric field $E$ is applied to a suspension of dispersed, non-spherical nanoparticles in contact with a thermal bath at a temperature $T$, it induces an electric dipole oriented along the particle’s major axis. The resulting electric torque tends to align the total dipole (permanent plus induced) with the field. Due to the inherent presence of thermal fluctuations, complete alignment is approached only for very intense electric fields with respect to the average strength of such fluctuations. The orientational (nematic) order parameter
\[
S(t)=\int_0^\pi \frac{1}{2}\!\left(3\cos^2\theta-1\right)p(\theta,t)\sin\theta\,\mathrm{d}\theta
\]
characterizes the degree of alignment, where $\theta$ is the angle between the external field and the particle symmetry axis, and $p(\theta,t)$ is the orientational probability density function. For disk-like particles (our main focus), $S$ saturates at $S_{\rm sat}=-\tfrac{1}{2}$ at high field amplitudes (perfect alignment with the particle diameter along the field). For convenience, we use the normalized quantity $\bar{S}\equiv S/S_{\rm sat}$, which varies from $0$ (random orientation) to $1$ (saturation).

Under a sudden change in the field amplitude, the system evolves from an initial value $\bar{S}_{\rm i}$ to a final one $\bar{S}_{\rm f}$ (Fig.~\ref{fig:fig1}). The orientational probability density $p(\theta,t)$ follows the Smoluchowski equation for rotational diffusion~\cite{felderhof2001kerr,felderhof2001nonlinear,jones2003rotational}:
\begin{equation}
    \begin{split}
    \frac{1}{D}\frac{\partial p(\theta,t)}{\partial t} = \mathcal{L}_{\xi,\sigma} p,\\
    \mathcal{L}_{\xi,\sigma}=
	\frac{1}{\sin\theta}\frac{\partial }{\partial \theta}\left(
	\sin\theta \ \frac{\partial}{\partial \theta}
	\right)
	+\\
	+\frac{1}{\sin\theta}
	\frac{\partial}{\partial \theta}
	\left(
    \xi\sin^2\theta  +
	2\sigma \sin^2\theta \cos\theta\right), 
\end{split}
\label{eq:smoluchowski_polar}
\end{equation}
where $D$ is the rotational diffusion coefficient and $\mathcal{L}_{\xi,\sigma}$ is the Smoluchowski operator.
The dimensionless coupling parameter $\xi=mE/(k_{\rm B}T)$ measures the permanent-dipole interaction energy $mE$, associated with the permanent-dipole moment $m$, in units of thermal energy $k_{\rm B}T$ ($k_{\rm B}$ is the Boltzmann constant). Meanwhile, $\sigma=\Delta\alpha^{e}E^2/(2k_{\rm B}T)$ is the corresponding dimensionless measure of the induced-dipole interaction energy $\Delta\alpha^{e}E^2/2$ associated with the anisotropic electric polarizability, $\Delta\alpha^{e}=\alpha_{\parallel}-\alpha_{\perp}$ (subscripts denote electric polarizability components parallel and perpendicular to the particle's symmetry axis, respectively).

The set of angular moments $f_\ell(t)\equiv \langle P_\ell(\cos\theta(t))\rangle$ (in particular, $f_2(t)=S(t)$) fully describes the relaxation and obeys (see Ref.~\cite{raikher_selective_2001} and SI for details)
\begin{equation}
\label{eq:fl_evol}
\frac{\mathrm{d}\mathbf{f}}{\mathrm{d}t}
= -D\big(\mathbf{M}_{\rm free} + \xi\,\mathbf{M}_{\rm perm} + 2\sigma\,\mathbf{M}_{\rm ind}\big)\mathbf{f}
+ D\,\mathbf{v}(\xi,\sigma),
\end{equation}
where $\mathbf{f}=(f_\ell)_{\ell\ge 1}$, $\mathbf{v}=\left(\tfrac{2}{5}\xi,\tfrac{4}{5}\sigma,0,\dots\right)$ and the matrices $\mathbf{M}_i$ are independent of $\xi$ and $\sigma$. Equivalently, solving the eigenvalue problem of $\mathcal{L}_{\xi,\sigma}$ yields the normal-modes representation
\begin{equation}
\label{eq:fl_tiempo}
f_\ell(t)=f_\ell(\infty)+\sum_n b_{n\ell}\,e^{-\lambda_n D\,(t-t_0)},
\end{equation}
with right/left eigenfunctions $\phi_n$/$\phi_n^\dagger$ and eigenvalues $\lambda_n$ of $\mathcal{L}$, and coefficients
$b_{n\ell}=\big(\langle \phi_n^\dagger|p(t_0)\rangle/\langle \phi_n^\dagger|\phi_n\rangle\big)\,\langle P_\ell|\phi_n\rangle$.
Since the transient speed is governed by $\lambda_n$, an optimal acceleration strategy should seek minimizing the weights $b_{n\ell}$ associated with the smallest eigenvalues~\cite{gal2020precooling}.

\eqref{eq:smoluchowski_polar}--\eqref{eq:fl_tiempo} hold for arbitrary temporal protocols $\xi(t)$ and $\sigma(t)$. In our experiments, we applied high-frequency AC electric fields, $E(t) = E_0\sin(2\pi\nu t)$, of different amplitudes $E_0$ to suppress non–electro-orientational effects (such as the presence of hydrodynamic torques \cite{shilov_polarization_2000}). Under these conditions, $\xi(t) = (mE_0/k_{\rm B}T) \sin(2\pi \nu t)$ and $\sigma(t) = (\Delta \alpha E_0^2/2k_{\rm B}T) \sin^2(2\pi \nu t)$. At sufficiently high frequency $\nu$, permanent dipoles cannot follow the rapid field oscillations and experience no net torque on average, whereas induced dipoles respond as under a DC field of the same RMS amplitude (see SI and Ref.~\cite{watanabe1984kerr}). Consequently, in what follows, we neglect the contribution of permanent dipoles to the orientational dynamics.

\section{Experimental observation of the Kovacs anomaly in nanoparticle electro-orientation}

\begin{figure*}[htpb]
    \centering
    \includegraphics[width=\linewidth]{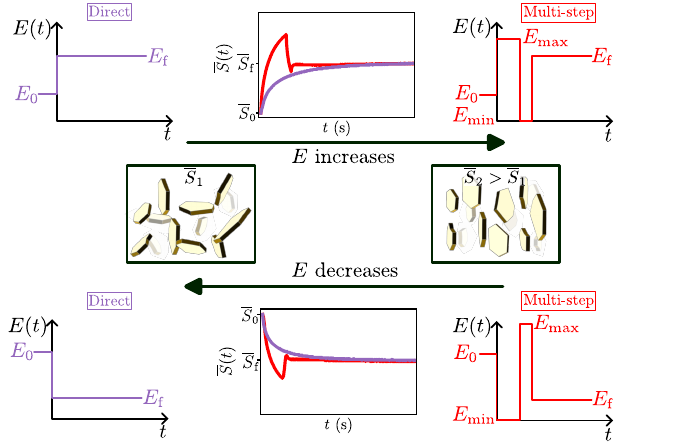}
    \caption{\textbf{Optimizing the alignment dynamics of Brownian particles in electric fields.} 
    Schematic representation of the experimental system in a poorly aligned (left) and a strongly aligned (right) configuration. 
    The upper panels show the time evolution of the electric field and the normalized order parameter $\bar{S}(t)$ during alignment processes (AP) using a direct protocol (purple) and a near time-optimal, three-step bang–bang protocol (red). 
    The lower panels depict the corresponding misalignment processes (MP). 
    In both cases, the three-step bang–bang protocol substantially reduces the response time compared with the direct one.
}
    \label{fig:fig1}
\end{figure*}

We consider two classes of processes defined by the relation between the initial and final values of the normalized order parameter, $\bar{S}_{\rm i}$ and $\bar{S}_{\rm f}$. When $\bar{S}_{\rm f}>\bar{S}_{\rm i}$ we refer to an alignment process (AP), whereas $\bar{S}_{\rm f}<\bar{S}_{\rm i}$ defines a misalignment process (MP). Among many possible choices, we focus here on two representative cases. In AP, the system always starts from a randomly oriented state ($\bar{S}_{\rm i}=0$), while in MP it departs from a state with $\bar{S}_{\rm i}=\bar{S}_{\max}=0.11$, attained at a maximum field amplitude $E_{\max}=5.88~\mathrm{V\,mm^{-1}}$. To realize these processes in practice, we employ two distinct families of electric-field \emph{protocols}. The simplest is a \emph{direct} protocol, consisting of a single step from the initial field $E_{\rm i}$ to the target field $E_{\rm f}$ at $t=0$ (Fig.~\ref{fig:fig1}, left). Alternatively, motivated by prior work in other implementations~\cite{patron2022thermal,patron2023non,patron2024minimum,prados2021optimizing}, we use multi-step (\textit{bang-bang}) protocols in which the intensity of the electric field is modulated between $E_{\max}$ and $E=0$ in a piecewise fashion until it is finally set to the target $E_{\rm f}$ (Fig.~\ref{fig:fig1}, right). The purpose of these protocols is to drive the system along a minimal-time (brachistochrone) trajectory in the space of orientational degrees of freedom, as will be made clear below. 

\begin{figure}[h]
    \centering
    \includegraphics[width=\linewidth]{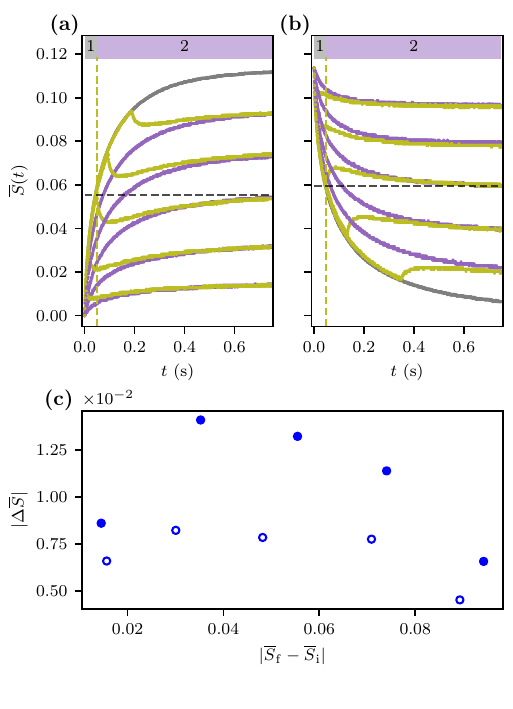}
    \caption{
    \textbf{Alignment (AP) and misalignment (MP) processes under direct and matched two-step protocols.} 
    \textit{(a) and (b):} Time evolution of the normalized orientational order parameter $\bar{S}(t)$ of NaMt nanoparticles immersed in an aqueous solution under direct (purple) and matched two-step (green) protocols for AP (a) and MP (b) at different final electric fields ($E_{\rm f}=1.96,\ 2.94,\ 3.91,\ 4.66,\ 5.29~\mathrm{V\,mm^{-1}}$). 
    Initial conditions are $E_{\rm i}=0$ for AP and $E_{\rm i}=5.88~\mathrm{V\,mm^{-1}}$ for MP. 
    Gray curves represent direct protocols to the limit fields, $E_{\rm i}\!\to\!E_{\max}=5.88~\mathrm{V\,mm^{-1}}$ (AP) and $E_{\rm i}\!\to\!E=0$ (MP). 
    As an example, the matched two-step protocol for the target value $\bar{S}_{\rm f}=0.055$ ($E_{\rm f}=3.91~\mathrm{V\,mm^{-1}}$; horizontal dashed line) is shown in the shaded area above each panel. 
    In this protocol, the system is first subjected to the limit field $E_{\max}$ (AP) or $E=0$ (MP) until the time instant $t_{\rm c}^{(\mathrm{K})}$ (vertical dashed line), where $\bar{S}(t_{\rm c}^{(\mathrm{K})})=\bar{S}_{\rm f}$, after which the field is switched to $E_{\rm f}$. 
    Error bars are smaller than the symbol size. 
    \textit{(c):} Amplitude of the Kovacs anomaly, $|\Delta\bar{S}|$, defined as the maximum deviation of $\bar{S}(t)$ from $\bar{S}_{\rm f}$ after $t_{\rm c}^{(\mathrm{K})}$, plotted as a function of the distance in $\bar{S}$ between the initial and final states, $|\bar{S}_{\rm f}-\bar{S}_{\rm i}|$. 
    Filled symbols correspond to AP and open symbols to MP.
    }
    \label{fig:fig2}    
\end{figure}

We first compare the direct and two-step protocols by tracking the time evolution of $\bar{S}(t)$ for both AP and MP of a NaMt aqueous suspension (see Fig.~\ref{fig:fig2} and Materials \& Methods). In the case of AP, the field is initially set to $E_{\max}$ for a time window $t_{\rm c}$ and then switched to different $E_{\rm f}$ values. We focus in particular on the \textit{matched} two-step protocol, originally introduced by Kovacs \emph{et~al.}~\cite{kovacs1979isobaric} to accelerate relaxation in glassy systems. In this protocol, $t_{\rm c}$ is chosen such that $\bar{S}(t_{\rm c})=\bar{S}_{\rm f}$; we denote this value by $t_{\rm c}^{(\mathrm{K})}$ (vertical dashed line in Fig.~\ref{fig:fig2}(a),(b) for the example $\bar{S}_{\rm f}=0.055$). The design of the matched protocol is rooted on the fact that the time scale for alignment is inversely proportional to the torque on the particles, which grows with the applied electric field. Therefore, one expects that initially applying an electric field higher than the target will speed-up the evolution, and the electric field can be switched to the target once the system has reached $\bar{S}_{\rm f}$.

Figure~\ref{fig:fig2}(a) illustrates the AP for several target states under direct and matched two-step protocols; for reference, the direct response to $E_{\max}$ is also shown. Under a direct protocol, $\bar{S}(t)$ increases monotonically toward $\bar{S}_{\rm f}$ (or $\overline{S}_{\max}$), marked with a horizontal dashed line for the example $\bar{S}_{\rm f}=0.055$. As expected, direct relaxation under $E_{\max}$ exhibits a faster transient than under any other $E_{\rm f}<E_{\max}$. In contrast, with the matched two-step protocol initially driven at $E=E_{\max}$, the trajectory displays a nonmonotonic transient after the switch to $E_{\rm f}$: an inverted shoulder emerges beyond $t_{\rm c}^{(\mathrm{K})}$. As a result, the curve rejoins the direct-protocol trajectory under $E_{\rm f}$ after the shoulder, and no overall speedup is achieved. 

The behavior for MP (Fig.~\ref{fig:fig2}(b)) is analogous. The system starts from $\bar{S}_{\max}$ and the direct protocols under $E=0$ and $E_{\rm f}$ are also shown for comparison. In the matched two-step protocol, the first time window sets $E=0$ until $\bar{S}$ reaches $\bar{S}_{\rm f}$, after which the field is switched to $E_{\rm f}$. As in AP, a Kovacs shoulder appears and the trajectory subsequently converges onto the direct-protocol curve, providing no reduction in the connection time.

All the cases considered and shown in Fig.~\ref{fig:fig2} share a common behavior. During the first window, the system follows the direct protocol under $E_{\max}$ (AP) or $E=0$ (MP). Upon switching to the second window, a Kovacs-type anomaly is consistently observed for all target states, indicating that the non-equilibrium pathway is governed by at least two degrees of freedom with distinct characteristic times, one exceeding $t_{\rm c}$. The figure further shows that the Kovacs shoulder is not equally pronounced for all targets, whose amplitude as a function of $|\bar{S}_{\rm f}-\bar{S}_{\rm i}|$ is depicted in Fig.~\ref{fig:fig2}(c) for both AP and MP. From the standpoint of acceleration, the least favorable cases are those with target states between the initial one and the degree of orientation that would be attained under a direct protocol under $E_{\max}$ (or to $E=0$ for MP). 

\begin{figure}[h]
    \centering
    \includegraphics[width=\linewidth]{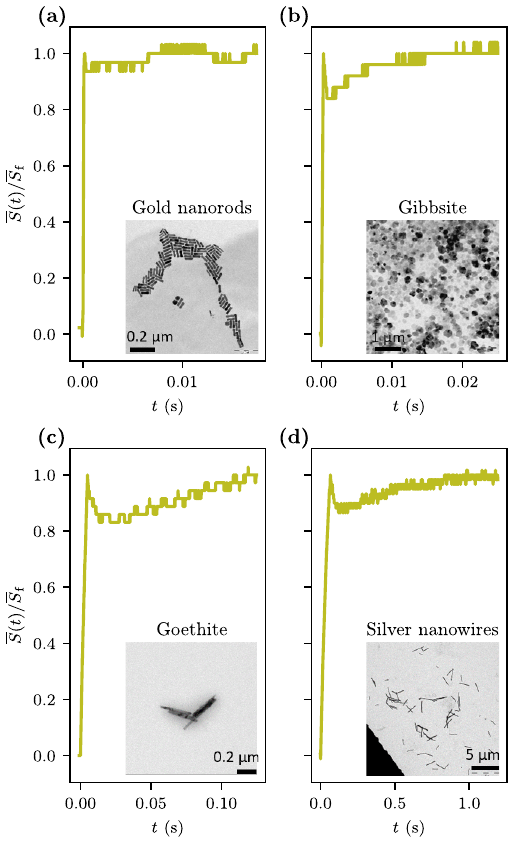}
    \caption{
    \textbf{Transient orientational dynamics for different particle systems under a matched two-step AP.} 
    Time evolution of the normalized order parameter $\bar{S}(t)/\bar{S}_{\rm f}$ for four representative materials, indicated in each panel: gold nanorods, gibbsite, goethite, and silver nanowires. 
    All measurements were performed under 100 kHz AC fields following alignment processes. 
    Insets show electron microscopy images of the corresponding samples, illustrating their morphology and characteristic sizes (scale bars as indicated). 
    Nearly monodisperse systems such as gold nanorods exhibit an almost single-exponential relaxation, while more polydisperse materials (e.g., silver nanowires) display a pronounced Kovacs shoulder.}
    \label{fig:fig3}    
\end{figure}

We next tested the generality of this phenomenon by examining the evolution under matched two-step protocols across systems with different properties—monodisperse and polydisperse, elongated and oblate, metallic and oxide—under the same high-frequency AC conditions (see Fig.~\ref{fig:fig3} and Materials \& Methods section). While not exhaustive, the comparison reveals a consistent trend: samples with narrower size distribution display only weak or barely discernible shoulders, whereas polydisperse suspensions exhibit a pronounced Kovacs feature. This behavior is consistent with two complementary considerations. First, a spread in particle sizes entails a spread in rotational diffusion coefficients, which naturally produces multiexponential relaxation~\cite{ArenasGuerrero2018b}. Second, \eqref{eq:fl_tiempo} shows through the the modal expansion of $S(t)$ that multiple intrinsic timescales may emerge even in monodisperse ensembles when both permanent and induced dipoles contribute. By contrast, when only induced dipoles are at play—as in our high-frequency AC regime—parity selection in the eigenmode expansion restricts the relaxation of $S(t)$ to even modes, leaving it dominated by the lowest nonzero eigenvalue of $\mathcal{L}$ that couples to $P_2$. The relaxation is therefore effectively single-exponential, and no Kovacs anomaly can arise (see SI for details). Together, these observations support our interpretation that the Kovacs shoulder originates from polydispersity under the present experimental conditions: nearly monodisperse gold nanorods  show a minimal effect, whereas a highly polydisperse sample (such as silver nanowires) displays a pronounced one.

\section{Optimization of the two-step protocol}

The above considerations are consistent with the established view that the Kovacs effect reflects the presence of multiple relaxation time scales~\cite{aquino2006kovacs,prados2010kovacs,peyrard2020memory,militaru2021kovacs}. In our case, the transient shoulder observed in $\bar{S}(t)$ after applying a matched two-step protocol (for $t\ge t_{\rm c}^{(\mathrm{K})}$) is accurately captured by a bi-exponential form,
\begin{equation}\label{eq:doble_exp}
    \bar{S}(t) \approx \bar{S}_{\rm f}
    + b_{\rm slow}\,e^{-(t-t_{\rm c})/\tau_{\rm slow}}
    + b_{\rm fast}\,e^{-(t-t_{\rm c})/\tau_{\rm fast}},
\end{equation}
where $t_{\rm c}$ denotes the switching time (equal to $t_{\rm c}^{(\mathrm{K})}$ for the matched protocol). The two contributions represent, respectively, particles that have not yet reached (\emph{slow} mode) or have overshot (\emph{fast} mode) the target degree of alignment (see SI for details). The amplitudes $b_{\rm slow}$ and $b_{\rm fast}$ have opposite signs, which breaks monotonicity and yields the Kovacs shoulder.

A closer inspection of the amplitudes in \eqref{eq:doble_exp} (which depend on $t_{\rm c}$; see SI) suggests a simple route to shape the post-switch dynamics through the duration of the first time window. Specifically, increasing $t_{\rm c}$ beyond the matched value $t_{\rm c}^{(\mathrm{K})}$ tends to suppress the slow contribution associated with the post-shoulder rise, thereby reducing the shoulder amplitude. Physically, this can be interpreted as giving the slowest subpopulation enough time to reach, on average, the target orientation before the switch. In the limiting case where $b_{\rm slow}$ becomes negligible, the relaxation for $t\ge t_{\rm c}$ is expected to be monotonic toward $S_{\rm f}$. Protocols that exploit this idea---attenuating the slowest decay mode to speed up the approach to the final state---have been proposed in related stochastic settings~\cite{gal2020precooling}.

\begin{figure}[h]
	\centering
	\includegraphics[width=\linewidth]{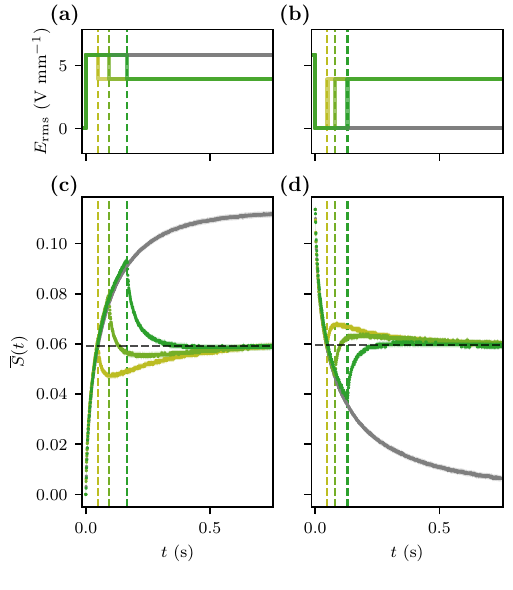}
	\caption{
    \textbf{Alignment (AP, left column) and misalignment (MP, right column) processes under different two-step protocols.} 
    \textit{(a) and (b)} RMS electric field, $E_{\mathrm{RMS}}(t)$, as a function of time for AP and MP. 
    Initial conditions are $E_{\rm i}=0$ for AP and $E_{\rm i}=5.88~\mathrm{V\,mm^{-1}}$ for MP, with a common final state corresponding to $E_{\rm f}=3.91~\mathrm{V\,mm^{-1}}$. 
    Three protocols are shown: a matched case ($t_{\rm c}=t_{\rm c}^{(\mathrm{K})}$), an improved case ($t_{\rm c}$ chosen so that $b_{\rm slow}\!\to\!0$), and an intermediate two-step protocol. 
    \textit{(c) and (d)} Time evolution of the normalized orientational order parameter $\bar{S}(t)$ under the protocols shown above. 
    The horizontal dashed line marks the target value $\bar{S}_{\rm f}=0.055$, and vertical dashed lines indicate the field-switching times. 
    Gray curves correspond to the direct protocols $E_{\rm i}\!\to\!E_{\max}=5.88~\mathrm{V\,mm^{-1}}$ (AP) and $E_{\rm i}\!\to\!E=0$ (MP), shown for comparison. 
    Error bars are smaller than the symbol size.}
	\label{fig:fig4}
\end{figure}

This strategy is experimentally tested in Fig.~\ref{fig:fig4}, which shows the time traces of $\bar{S}(t)$ (Figs.~\ref{fig:fig4}(c),(d) for AP and MP, respectively) for several two-step protocols differing in $t_{\rm c}$ (Figs.~\ref{fig:fig4}(a),(b)), at fixed $E_{\rm f}=3.91~\mathrm{V\,mm^{-1}}$ (yielding $\bar{S}_{\rm f}=0.055$). We highlight the outcome of an \emph{improved two-step protocol}, defined by the minimal $t_{\rm c}$ that removes the hump in $\bar{S}(t)$; in this case, the subsequent evolution is strictly monotonic. Somewhat counterintuitively, the approach to the final state is accelerated solely by increasing the duration of the first window.

\begin{figure}[h]
	\centering
	\includegraphics[width=0.45\textwidth]{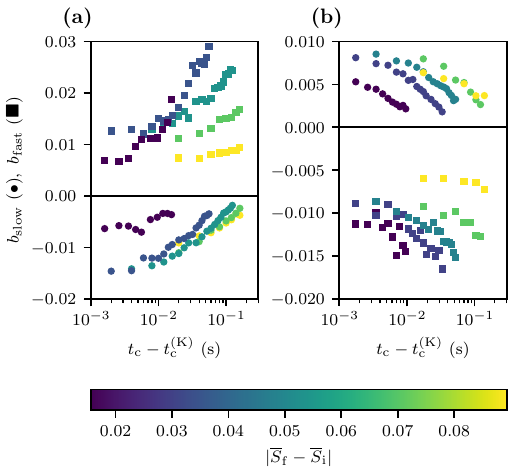}
	\caption{
    \textbf{Evolution of the fast and slow mode amplitudes with protocol duration.} 
    Amplitudes of the fast (squares) and slow (circles) decay modes as a function of the difference between the duration of the first time window, $t_{\rm c}$, and its matched value, $t_{\rm c}^{(\mathrm{K})}$, for AP (panel (a)) and MP (panel (b)). 
    Colors encode the distance in the order parameter between the initial and final steady states, $|\bar{S}_{\rm f}-\bar{S}_{\rm i}|$, as indicated in the color bar below. 
    Error bars are smaller than the symbol size.}
	\label{fig:fig5}
\end{figure}

To quantify these trends, we fit $\bar{S}(t\ge t_{\rm c})$ to \eqref{eq:doble_exp}. The slow and fast modes amplitudes are depicted in Figs.~\ref{fig:fig5}(a),(b) for AP and MP as functions of $t_{\rm c}-t_{\rm c}^{(\mathrm{K})}$ for different $\overline{S}_{\rm f}$ values. Rather than labeling by $\bar{S}_{\rm f}$, we group the data by the distance to the initial state, $|\bar{S}_{\rm f}-\bar{S}_{\rm i}|$. The analysis confirms that $b_{\rm slow}$ decreases to zero as $t_{\rm c}$ increases, while $b_{\rm fast}$ concomitantly grows, consistent with the suppression of the slowest mode and the recovery of monotonic relaxation.

\section{From bang--bang theory to the design of an optimal control protocol}

We seek to minimize the time required to connect two steady states of the orientational dynamics, subject to the evolution law given by \eqref{eq:fl_evol} applied to each size subpopulation and neglecting permanent-dipole contributions. In this regime, the sole control parameter is $\sigma(t)\propto E(t)^2$, which is bounded by the capabilities of the signal generator, $0\le E(t)\le E_{\max}$. 

For minimum-time problems of this type (bounded input and dynamics linear in the control), theory predicts that the time-optimal solution is of the bang--bang class, with as many switches between the extremal control values as there are independent dynamical degrees of freedom~\cite{chen2010fast,ding2020smooth,prados2021optimizing,ruiz2022optimal,patron2022thermal,patron2024minimum,guery2023driving}. In our case, each monodisperse subpopulation relaxes effectively with a single time scale, so the corresponding optimal protocol is prescribed by a matched two-step policy \cite{patron2022thermal}. For a polydisperse suspension, however, the number of independent time scales is, in principle, unbounded; an exactly optimal bang--bang sequence would thus require infinitely many switches and is not experimentally practical.

\begin{figure}[htpb]
	\centering
	\includegraphics[width=\linewidth]{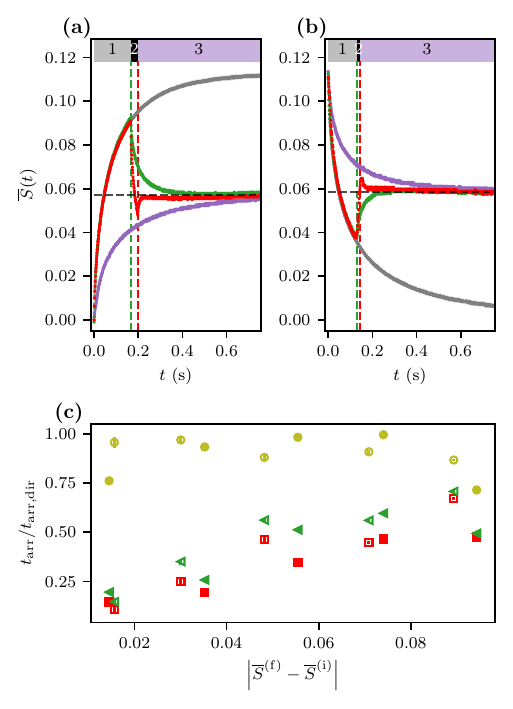}
	\caption{
    \textbf{Alignment (AP) and misalignment (MP) processes under our near-optimal protocol.} 
    (a) Time evolution of the normalized orientational order parameter $\bar{S}(t)$ under direct (purple), improved two-step (green), and near-optimal (red) protocols for AP to a final degree of alignment $\bar{S}_{\rm f}=0.055$ ($E_{\rm f}=3.91~\mathrm{V\,mm^{-1}}$; horizontal dashed line). 
    Initial conditions are $E_{\rm i}=0$ for AP and $E_{\rm i}=5.88~\mathrm{V\,mm^{-1}}$ for MP. Gray curves represent direct protocols to the limit fields, $E_{\rm i}\!\to\!E_{\max}=5.88~\mathrm{V\,mm^{-1}}$ (AP) and $E_{\rm i}\!\to\!E=0$ (MP). 
    Our near-optimal protocol is illustrated in the shaded area above each panel: the system is first subjected to the limit field $E_{\max}$ (AP) or $E=0$ (MP), then to the opposite extreme, and finally to $E_{\rm f}$. Vertical dashed lines indicate the field-switching times. Error bars are smaller than the symbol size. (b) Same as (a), but for MP. 
    (c) Relaxation times as a function of the distance between the initial and final states, $|\bar{S}_{\rm f}-\bar{S}_{\rm i}|$, for AP (filled markers) and MP (open markers). 
    The relaxation time of the near-optimal protocol (squares) corresponds to the prescribed duration set by the user. 
    For comparison, the relaxation times of the direct (dots) and improved two-step (triangles) protocols are defined as the first instant when the experimental curve reaches $\bar{S}_{\rm f}$ (see Methods). 
    Error bars represent the temporal precision of the oscilloscope. 
    All relaxation times are normalized by that of the direct protocol.}
	\label{fig:fig6}
\end{figure}

We therefore propose an experimentally feasible and near-optimal protocol that is easy to implement yet highly effective. As shown in Figs.~\ref{fig:fig6}(a),(b) for both AP and MP, we apply a three-step protocol schematically represented in the upper shaded region of each plot: starting from $E_{\max}$ (AP) or $E=0$ (MP), each labelled with ``1'', we switch once to the opposite extreme (labelled with ``2'') and finally to the target field $E_{\rm f}$ (labelled with ``3''). The durations of the first two windows are chosen by leveraging the benefits of the improved two-step protocol: each window is adjusted to the minimal $t_{\rm c}$ that nullifies the slow mode (i.e., $b_{\rm slow}\to 0$) in the bi-exponential relaxation, so that the subsequent evolution is monotonic. For reference, we also include the trajectories produced by the direct and improved two-step protocols. The guiding principle is that removing the slowest mode at each stage produces an exponential speedup of the remaining dynamics.

The effectiveness of our optimal design is quantified in Fig.~\ref{fig:fig6}(c), which reports the arrival time $t_{\rm arr}$ needed to reach the target degree of alignment as a function of $|\bar{S}_{\rm f}-\bar{S}_{\rm i}|$ (see Materials \& Methods for details on its estimation). For comparison, we also show $t_{\rm arr}$ for the matched and improved two-step protocols, for both AP and MP. To emphasize gains or losses relative to simple relaxation, $t_{\rm arr}$ is normalized by that of the direct protocol at $E_{\rm f}$, $t_{\rm arr,dir}$.

\section{Discussion}

Electro-optic birefringence in field-oriented nanoparticle suspensions exhibits a Kovacs-type anomaly, signaling the presence of multiple exponential decay modes in the relaxation pathway. The dynamics differs qualitatively across the stages of a two-step protocol. During the first window of AP, electric torques dominate over Brownian fluctuations and drive particles toward the minima of the field-induced potential, at rates that depend on their instantaneous orientations and characteristic time scales. After the switch, two antagonistic contributions coexist: particles that were already close to the extreme state partially misalign because Brownian torques gain relative weight, whereas particles still en route toward alignment relax more slowly under the reduced field. Their superposition generates a transient nonmonotonic shoulder in $\bar{S}(t)$ before the trajectory converges to the new state.

The anomaly also displays the hallmark features of memory effects: its magnitude and even its sign depend sensitively on the initial and target states. When $\bar{S}_{\rm f}$ lies close to $\bar{S}_{\rm i}$, the first window is too short to significantly advance either the fastest or the slowest subpopulations, and the shoulder is weak. Conversely, as $\bar{S}_{\rm f}$ approaches the limit states reached under direct protocols—$E_{\max}$ for AP or $E=0$ for MP—most particles have already attained the corresponding degree of alignment by the end of the first window, and the subsequent evolution is predominantly monotonic. An asymmetry between AP and MP is naturally expected because relaxation time scales depend on field strength: AP typically proceeds from a fast initial transient (at large field) to a slower approach after the switch, yielding more pronounced overshoots, whereas MP shows the inverse pattern and correspondingly smaller shoulders. Similar directional asymmetries have also been observed in other systems when steering the dynamics in opposite directions~\cite{ibanez2024heating}.

The modal expansion shows that a monodisperse system can also exhibit multiscale dynamics when permanent and induced dipoles are simultaneously active: parity is then broken and the lowest eigenmode contributes to the relaxation of each angular moment $f_\ell$. These findings reconcile the general mechanism underlying the Kovacs effect—the attempt to steer many degrees of freedom and characteristic times with a single control parameter—with its specific manifestation here. They reveal two complementary sources of multiscale relaxation: an intrinsic one, arising from the hierarchy of eigenmodes governing each subpopulation, and an extrinsic one, due to the coexistence of many subpopulations with distinct rotational diffusion coefficients. Even a monodisperse system thus remains inherently multiscale whenever the chosen observable is not itself an eigenfunction of the Smoluchowski operator. Nevertheless, our experiments were conducted under high-frequency AC fields, a regime in which permanent dipoles cannot follow the oscillations and thus suffer no net torque; the orientational dynamics is governed by induced dipoles. Under these conditions, the Kovacs shoulder arises exclusively from polydispersity: a distribution of particle sizes leads to a spectrum of rotational diffusion coefficients and, consequently, to multiscale relaxation. 

In the search for minimal-time steering, our results show that the matched two-step (Kovacs) protocol does not outperform the direct protocol, irrespective of the target state (see Fig.~\ref{fig:fig6}(c)). By contrast, the improved two-step strategy—choosing the first-window duration to suppress the slowest mode ($b_{\rm slow}\!\to\!0$)—restores monotonic relaxation and accelerates the approach to the target state (Figs.~\ref{fig:fig4} and \ref{fig:fig6}(c)). Building on this idea, we proposed a simple three-step design that alternates the field extremes once before setting $E_{\rm f}$. Each window is tuned using the improved two-step criterion so that the slowest active mode is eliminated stage by stage. This near-optimal construction yields a further, albeit modest, reduction in relaxation time and stabilizes the post-switch evolution of $\bar{S}(t)$ (Fig.~\ref{fig:fig6}). The relative gain is largest when $|\bar{S}_{\rm f}-\bar{S}_{\rm i}|$ is small, where the direct trajectory has a steep initial slope and can be most effectively accelerated; for widely separated states, the dynamics necessarily tracks much of the direct path governed by the system’s intrinsic time scales.

Strictly speaking, the time-optimal bang--bang solution for a polydisperse ensemble would require as many switches as independent time scales—formally, an unbounded number—and is therefore impractical. Our approach embraces this constraint and targets the dominant physics: the first window advances the entire population until the slowest subpopulation reaches the target on average; the second window allows the overshot fraction to relax back without significantly erasing the progress of the slowest group; residual evolution is then governed by intermediate modes and is comparatively minor. Additional switching could further reduce this residual, but with diminishing returns and increased implementation complexity. In practice, the number of switches should be chosen according to the desired precision (tolerance on $|\bar{S}(t)-\bar{S}_{\rm f}|$) and experimental uncertainty, balancing marginal speedups against control overhead.

\subsection*{Materials}

Experiments were mainly performed using suspensions of sodium montmorillonite (NaMt) platelets in ultrapure water containing NaCl at a concentration of $1\,\mathrm{g/L}$. NaMt particles were obtained by sodium homoionization of bentonite (Sigma--Aldrich, USA). The resulting particles are lamellar platelets, with negatively charged faces and a pH-dependent edge charge, and were characterized by environmental scanning electron microscopy. Their average diameter is $1.7\pm0.6\,\mu\mathrm{m}$~\cite{ArenasGuerrero2016}.

For comparison, we also investigated the presence of Kovacs anomaly in aqueous suspensions of gibbsite platelets, goethite needles, gold nanorods and silver nanowires (see Figure~\ref{fig:fig3} and Table~\ref{tab:size} for mean sizes and standard deviations). Gibbsite and goethite are oxide/hydroxide colloids whose surface charge density can be spatially heterogeneous, which may lead to a permanent electric dipole moment~\cite{Rica2009,Ahualli2017}. In contrast, the surface charge of gold nanorods and silver nanowires is mainly determined by their stabilizing layers---positively charged cetyltrimethylammonium bromide (CTAB) for gold nanorods and negatively charged polyvinylpyrrolidone (PVP) for silver nanowires~\cite{ArenasGuerrero2019,ArenasGuerrero2021}. In addition, due to their metallic character, gold and silver particles are expected to exhibit a strong induced dipole under the applied electric field.

\begin{table}[hbtp] 
    \centering \resizebox{\linewidth}{!}{ 
    \begin{tabular}{|c|c|c|} 
    \hline Material & Size & Standard Deviation 
    \\ \hline\hline 
    Gibbsite & Diameter: 257 nm & 40 nm 
    \\ 
    ($\gamma-$Al(OH)$_3$, sinthetic, \cite{Ahualli2017}) & & 
    \\ \hline 
    Goethite needles & Length: 460 nm & 120 nm 
    \\ 
    (FeOOH, Sigma-Aldrich, USA) & & 
    \\ \hline Gold nanorods & Length: 56 nm & 9 nm 
    \\
    (synthetic, \cite{ArenasGuerrero2019}) & & 
    \\ \hline 
    Silver nanowires & Length: 2000 nm & 900 nm 
    \\ (commercial, Plasma Chem Laboratory, Germany) & & 
    \\ \hline 
    \end{tabular} 
    } 
    \caption{Characteristic size (mean) and polydispersity (standard deviation) of the samples in Fig.~\ref{fig:fig3}.}
    \label{tab:size} 
\end{table}

\subsection*{Experimental methods}

Partial alignment of a suspension of non-spherical particles renders the medium optically anisotropic. In particular, an electric-field–induced birefringence $\Delta n=n_{\parallel}-n_{\perp}$ arises, where $n_{\parallel}$ and $n_{\perp}$ are the refractive indices parallel and perpendicular to the field direction, respectively. It is well established that $\Delta n$ is proportional to the orientational order parameter~\cite{Fredericq1973},
\begin{equation}
    \Delta n(t) \;=\; \Delta n_{\max}\,\bar S(t),
\end{equation}
where $\Delta n_{\max}$ denotes the asymptotic birefringence at large fields (all particles with their major axis along the field) and $\bar S(t)=S(t)/S_{\mathrm{sat}}$ with $S_{\mathrm{sat}}=-0.5$. For a 2\,g/L NaMt dispersion, we find $\Delta n_{\max}=4.68(6)\times 10^{-6}$ (see SI for details).

Electric birefringence was measured with a custom-built optical setup in which a laser beam traverses the sample and polarization optics; the transmitted intensity was detected by a photodiode and recorded as a function of time (further details in SI and Ref.~\cite{piazza1986matrix}). The applied electric field was sinusoidal at 100\,kHz with amplitude $E_0$ and root-mean-square value $E_{\mathrm{RMS}}=E_0/\sqrt{2}$. In all experiments, the field strength was kept below $E_{\mathrm{RMS}}=E_{\max}\simeq5.9~\mathrm{V/mm}$, which corresponds to the maximum applied value throughout this work.

\subsection*{Estimating the arrival time}

Based on the results in Fig.~\ref{fig:fig6}, after implementing the near-optimal protocol we consider that the system reaches the target state at the user-prescribed time $t_{\mathrm{opt}}$, given by the sum of the durations of the two windows. Although the state continues to evolve slightly thereafter (the second window follows the same criterion as in the improved two-step protocol), the subsequent excursion of $\bar{S}(t)$ around $\bar{S}_{\mathrm{f}}$ is small. We therefore construct a tolerance band for $\bar{S}_{\mathrm{f}}$ whose half-width equals the standard deviation of $\bar{S}(t\ge t_{\rm arr, opt})$ under the optimal protocol. Once this tolerance band is defined, the arrival time $t_{\rm arr}$ for any other protocol is estimated as the first time $t$ at which the corresponding experimental trace $\bar{S}(t)$ enters the band and remains within it. The same tolerance band is used for all protocols that share the same initial and target states.

\subsection*{Acknowledgements}
This work was supported by grants PID2021-127427NB-I00 and PID2024-161166NB-I00 funded by MICIU/AEI/10.13039/501100011033 and by ERDF, UE. M.I. acknowledges support from an FPU fellowship (FPU21/02569) granted by Ministerio de Ciencia, Innovaci\'on y Universidades (Spain).

%

\end{document}